\documentclass{article}
\usepackage{times}
\usepackage{amsfonts}
\usepackage{graphicx}
\begin{document}
\noindent
{\Large SCHROEDINGER vs. NAVIER-STOKES}\\
\vskip1cm
\noindent
{\bf P. Fern\'andez de C\'ordoba}$^{a}$, {\bf  J.M. Isidro}$^{b}$ and {\bf J. Vazquez Molina}$^{c}$\\
Instituto Universitario de Matem\'atica Pura y Aplicada,\\ Universidad Polit\'ecnica de Valencia, Valencia 46022, Spain\\
${}^{a}${\tt pfernandez@mat.upv.es}, ${}^{b}${\tt joissan@mat.upv.es},\\
${}^{c}${\tt joavzmo@etsii.upv.es}\\
\vskip.5cm
\noindent
{\bf Abstract} Quantum mechanics has been argued to be a coarse--graining of some underlying deterministic theory. Here we support this view by establishing a map between certain solutions of the Schroedinger equation, and the corresponding solutions of the irrotational Navier--Stokes equation for viscous fluid flow. As a physical model for the fluid itself we propose the quantum probability fluid. It turns out that the (state--dependent) viscosity of this fluid is proportional to Planck's constant, while the volume density of entropy is proportional to Boltzmann's constant. Stationary states have zero viscosity and a vanishing time rate of entropy density. On the other hand, the nonzero viscosity of nonstationary states provides an information--loss mechanism whereby a deterministic theory (a classical fluid governed by the Navier--Stokes equation) gives rise to an emergent theory (a quantum particle governed by the Schroedinger equation).

\tableofcontents

\section{Introduction}\label{dyctus}

Interaction with an environment provides a mechanism whereby classical behaviour can emerge from a quantum system \cite{ZUREK}. At the same time, however, dissipation into an environment can change this picture towards the opposite conclusion. Indeed certain forms of quantum behaviour have been {\it experimentally}\/ shown to arise within classical systems subject to dissipation \cite{COUDER, DISIPACION}. Now systems in thermal equilibrium are well described by classical thermostatics, while small deviations from thermal equilibrium can be described by the classical thermodynamics of irreversible processes \cite{ONSAGER}. It is sometimes possible to model long--wavelength dissipative processes through the dynamics of {\it viscous}\/ fluids. Fluid viscosity provides a relatively simple dissipative mechanism, a first deviation from ideal, frictionless behaviour. Two relevant physical quantities useful to characterise viscous fluids  are shear viscosity $\eta$ and the entropy per unit 3--volume, $s$ \cite{LANDAUFLUID}. In a turn of events leading back to the Maldacena conjecture \cite{MALDACENA} it was found that, for a wide class of thermal quantum field theories in 4 dimensions, the ratio $\eta/s$ for the quark--gluon plasma must satisfy the inequality \cite{PLASMA}
\begin{equation}
\frac{\eta}{s}\geq\frac{\hbar}{4\pi k_B}.
\label{ueber}
\end{equation}
The predicted value of the ratio $\eta/s$ for the quark--gluon plasma has found experimental confirmation \cite{QCD}. The simultaneous presence of Planck's constant $\hbar$ and Boltzmann's constant $k_B$ reminds us that we are dealing with theories that are both {\it quantum}\/ and {\it thermal}\/. 

One might be inclined to believe that these two properties, {\it quantum}\/ on the one hand, and {\it thermal}\/ on the other, are separate. One of the purposes of this paper is to show that this predisposition must be modified, at least partially, because the terms {\it quantum}\/ and {\it thermal}\/ are to a large extent linked (see {\it e.g.}  \cite{NOI1, PADDY} and refs. therein). In fact, that these two properties belong together follows from the analysis of  refs. \cite{DISIPACION, ZUREK}, even if the conclusions of these two papers seem to point in opposite directions.

In this article we elaborate on a theoretical framework that can accomodate the ideas of the previous paragraph. In plain words, this framework can be summarised in the statement  {\it quantum $=$ classical $+$ dissipation}\/, although of course this somewhat imprecise sentence must be made precise. To begin with, we will restrict our analysis to quantum systems with a finite number of degrees of freedom. So we will be dealing not with theories of fields, strings and branes, but with plain quantum mechanics instead.  

In the early days of quantum mechanics, Madelung  provided a very intuitive physical interpretation of the Schroedinger wave equation in terms of a probability fluid  \cite{MADELUNG}. Decomposing the complex wavefunction $\psi$ into amplitude and phase, Madelung transformed the Schroedinger wave equation into an equivalent set of two: the quantum Hamilton--Jacobi equation, and the continuity equation. Further taking the gradient of the phase of $\psi$, Madelung arrived at a velocity field satisfying the Euler equations for an {\it ideal}\/ fluid. In Madelung's analysis, the quantum potential $U$ is  interpreted as being (proportional to) the pressure field within the fluid. It is important to stress that Madelung's fluid was ideal, that is, {\it frictionless}\/. Independently of this analogy, Bohm suggested regarding the quantum potential $U$ as a force field that the quantum particle was subject to, in addition to any external, classical potential $V$ that might also be present \cite{BOHM}.

There exists yet a third, so far unexplored alternative to Madelung's and Bohm's independent interpretations of the quantum potential. In this alternative, explored here, {\it the quantum potential is made to account for a dissipative term in the equations of motion of the probability fluid}\/. The velocity field no longer satisfies Euler's equation for an ideal fluid---instead it satisfies the Navier--Stokes equation for a {\it viscous}\/ fluid. It is with this viscosity term in the Navier--Stokes equation, and its physical interpretation as deriving from the Schroedinger equation, that we will be concerned with in this paper.

It has long been argued that quantum mechanics must emerge from an underlying classical, deterministic theory via some coarse--graining, or information--loss mechanism \cite{ELZE1, ELZE2, FINSTER1, GALLEGO1, GALLEGO2, THOOFT0, THOOFT1, THOOFT2}; one refers to this fact as the {\it emergence property}\/ of quantum mechanics \cite{CARROLL}. Many emergent physical theories admit a thermodynamical reformulation, general relativity being perhaps the best example \cite{PADMANABHAN, VERLINDE}. Quantum mechanics is no exception \cite{DEBROGLIE, MATONE}; in fact our own approach \cite{NOI1, NOI2} to the emergence property of quantum mechanics exploits a neat correspondence with the classical thermodynamics of irreversible processes \cite{ONSAGER}. 

In this article, the dissipation that is intrinsic to the quantum description of the world will be shown to be ascribable to the viscosity $\eta$ of the {\it quantum probability fluid}\/ whose density equals Born's amplitude squared $\vert\psi\vert^2$. Moreover, the viscosity $\eta$ will turn out to be proportional to $\hbar$, thus vanishing in the limit $\hbar\to 0$. Now mechanical action (resp. entropy) is quantised in units of Planck's constant $\hbar$ (resp. Boltzmann's constant $k_B$),  and Eq. (\ref{ueber}) contains these two quanta. (Concerning Boltzmann's constant $k_B$ as a quantum of entropy, see refs. \cite{LANDAUER, VERLINDE}). Hence an important implication of our statement  {\it quantum $=$ classical $+$ dissipation}\/ is that quantum and thermal effects are inextricably linked.

Some remarks on conventions are in order; we follow ref. \cite{LANDAUFLUID}. The viscosity properties of a fluid can be encapsulated in the viscous stress tensor  $\sigma'_{ik}$,
\begin{equation}
\sigma'_{ik}:=\eta\left(\frac{\partial v_i}{\partial x_k}+\frac{\partial v_k}{\partial x_i}-\frac{2}{3}\delta_{ik}\frac{\partial v_l}{\partial x_l}\right)+\zeta\delta_{ik}\frac{\partial v_l}{\partial x_l},
\label{sorten}
\end{equation}
where $\eta$ (shear viscosity) and $\zeta$ (bulk viscosity) are positive coefficients, and the $v_i$ are the components of the velocity field ${\bf v}$ within the fluid. Then the Navier--Stokes equation reads 
\begin{equation}
\frac{\partial{\bf v}}{\partial t}+\left({\bf v}\cdot\nabla\right){\bf v}+\frac{1}{\rho}\nabla p-\frac{\eta}{\rho}\nabla^2{\bf v}-\frac{1}{\rho}\left(\zeta+\frac{\eta}{3}\right)\nabla\left(\nabla\cdot{\bf v}\right)=0.
\label{nabiq}
\end{equation}
Here $p$ is the pressure, and $\rho$ the density of the fluid. In the particular case of {\it irrotational}\/ flow considered here,  the Navier--Stokes equation simplifies to
\begin{equation}
\frac{\partial{\bf v}}{\partial t}+\left({\bf v}\cdot\nabla\right){\bf v}+\frac{1}{\rho}\nabla p-\frac{\eta'}{\rho}\nabla^2{\bf v}=0, \qquad \eta':=\zeta+\frac{4\eta}{3}.
\label{noxba}
\end{equation}
{}For notational simplicity, in what follows we will systematically write $\eta$ for the viscosity coefficient $\eta'$ just defined, bearing in mind, however, that we will always be dealing with Eq. (\ref{noxba}) instead of (\ref{nabiq}).

The above must be supplemented with the continuity equation and the equation for heat flow. If $T$ denotes the temperature and $\kappa$ the thermal conductivity of the fluid, then the equation governing heat transfer within the fluid reads
\begin{equation}
\rho T\left(\frac{\partial s}{\partial t}+({\bf v}\cdot\nabla)s\right)-\sigma_{ik}'\frac{\partial v_i}{\partial x_k}-\nabla\cdot\left(\kappa\nabla T\right)=0.
\label{calor}
\end{equation}

We will use the notations ${\cal I}$ and ${\cal S}$ for mechanical action and entropy, respectively, while the dimesionless ratios ${\cal I}/\hbar$ and ${\cal S}/2k_B$ will be denoted in italic type:
\begin{equation}
I:=\frac{{\cal I}}{\hbar}, \qquad S:=\frac{{\cal S}}{2k_B}.
\label{noetattio}
\end{equation}
The factor of $2$ multiplying $k_B$, although conventional, can be justified. By Boltzmann's principle, the entropy of a state is directly proportional to the logarithm of the probability of that state. In turn, this is equivalent to Born's rule:
\begin{equation}
{\rm (Boltzmann)}\quad{\cal S}=k_B\ln\left({\Big\vert}\frac{\psi}{\psi_0}{\Big\vert}^2\right)\Longleftrightarrow \vert\psi\vert^2=\vert\psi_0\vert^2\exp\left(\frac{{\cal S}}{k_B}\right)\quad{\rm (Born)}.
\label{volman}
\end{equation}
Above, $\vert\psi_0\vert$ is the amplitude of a fiducial state $\psi_0$ with vanishing entropy.   Such a fiducial state is indispensable because the argument of the logarithm in Boltzmann's formula must be dimensionless. It is convenient to think of  $\psi_0$ as being related to a 3--dimensional length scale $l$ defined through
\begin{equation}
l:=\vert\psi_0\vert^{-2/3}.
\label{skala}
\end{equation}
One can also think of $\psi_0$ as a normalisation factor for the wavefunction.

\section{The physics of Navier--Stokes from Schroedinger}\label{vscqeff}

\subsection{Computation of the viscosity}\label{fluirro}

Our starting point is Madelung's rewriting of the Schroedinger equation for a mass $m$ subject to a static potential $V=V({\bf x})$,
\begin{equation}
{\rm i}\hbar\frac{\partial\psi}{\partial t}+\frac{\hbar^2}{2m}\nabla^2\psi-V\psi=0,
\label{recontra}
\end{equation}
by means of the substitution
\begin{equation}
\psi=\psi_0\exp\left(S+\frac{{\rm i}}{\hbar}{\cal I}\right)=\psi_0A\exp\left(\frac{{\rm i}}{\hbar}{\cal I}\right), \qquad A:={\rm e}^S.
\label{ansaz}
\end{equation}
This produces, away from the zeroes of $\psi$, an equation  whose imaginary part is the continuity equation for the quantum probability fluid,
\begin{equation}
\frac{\partial S}{\partial t}+\frac{1}{m}\nabla S\cdot\nabla {\cal I}+\frac{1}{2m}\nabla^2{\cal I}=0,
\label{konntt}
\end{equation}
and whose real part is the quantum Hamilton--Jacobi equation:
\begin{equation}
\frac{\partial {\cal I}}{\partial t}+\frac{1}{2m}(\nabla {\cal I})^2+V+U=0.
\label{arel}
\end{equation}
Here
\begin{equation}
U:=-\frac{\hbar^2}{2m}\frac{\nabla^2 A}{A}=-\frac{\hbar^2}{2m}\left[\left(\nabla S\right)^2+\nabla^2S\right]
\label{cupot}
\end{equation}
is the quantum potential \cite{BOHM}. Next one defines the velocity field of the quantum probability fluid
\begin{equation}
{\bf v}:=\frac{1}{m}\nabla {\cal I}.
\label{atled}
\end{equation}
Then the gradient of Eq. (\ref{arel}) equals
\begin{equation}
\frac{\partial{\bf v}}{\partial t}+\left({\bf v}\cdot\nabla\right){\bf v}+\frac{1}{m}\nabla U+\frac{1}{m}\nabla V=0.
\label{kblm}
\end{equation}
The flow (\ref{atled}) is irrotational. We will sometimes (though not always) make the assumption of incompressibility, $\nabla\cdot{\bf v}=0$. This reduces to the requirement that the phase ${\cal I}$ satisfy the Laplace equation,
\begin{equation}
\nabla^2{\cal I}=0.
\label{laplas}
\end{equation}
We will see in Eq. (\ref{nouec}) that the above Laplace equation is an equivalent restatement of the semiclassicality condition.

At this point we deviate from Madelung's reasoning and compare Eq. (\ref{kblm}) not to Euler's equation for an ideal fluid, but to the Navier--Stokes equation instead, Eq. (\ref{noxba}). For the correspondence to hold, we first identify $(\nabla p)/\rho$ with $(\nabla V)/m$. Second, it must hold that
\begin{equation}
\frac{1}{m}\nabla U+\frac{\eta}{\rho}\nabla^2{\bf v}=0.
\label{atlee}
\end{equation}
That is, the gradient of the quantum potential must exactly compensate the viscosity term in the fluid's equations of motion. Thus frictional forces within the fluid are quantum in nature. Altogether, we have established the following:

{\bf Theorem 1} {\sl Whenever condition (\ref{atlee}) holds, the gradient of the quantum Hamilton--Jacobi equation, as given by Eq. (\ref{kblm}), is a Navier--Stokes equation for irrotational, viscous flow:
\begin{equation}
\frac{\partial{\bf v}}{\partial t}+\left({\bf v}\cdot\nabla\right){\bf v}-\frac{\eta}{\rho}\nabla^2{\bf v}+\frac{1}{\rho}\nabla p=0.
\label{reger}
\end{equation}
Here the pressure $p$ of the quantum probability fluid and the mechanical potential $V$ are related as per 
\begin{equation}
\frac{1}{\rho}\nabla p=\frac{1}{m}\nabla V,
\label{potpre}
\end{equation}
while the density $\rho$ of the fluid is given by
\begin{equation}
\rho=m\vert\psi\vert^2=\frac{m}{l^3}{\rm e}^{2S}=\frac{m}{l^3}A^2.
\label{felicita}
\end{equation}}

Given $V$, $m$ and $\rho$, the equation $(\nabla p)/\rho=(\nabla V)/m$ defines a vector field ${\bf p}=\rho\nabla V/m$, that however need not be a gradient field $\nabla p$. We will see later (theorem 4) that, at least in the classical limit, the above equation is integrable, thus defining a scalar function $p$ such that ${\bf p}=\nabla p$.

The order of magnitude of the viscosity coefficient $\eta$ can be inferred from Eqs. (\ref{cupot}), (\ref{atled}) and (\ref{atlee}): since $U$ is $O(\hbar^2)$ and ${\cal I}$ is $O(\hbar)$, we conclude:

{\bf Theorem 2} {\sl Whenever condition (\ref{atlee}) holds, the viscosity coefficient $\eta$ of the quantum probability fluid is proportional to Planck's constant:}
\begin{equation}
\eta=\frac{1}{l^3}O\left(\hbar\right).
\label{ordehache}
\end{equation}

It is worthwhile stressing that Eq. (\ref{ordehache}) only provides an order of magnitude for $\eta$ as a function of $\hbar$---namely, $\eta$ is a linear function of $\hbar$. The denominator $l^3$ has been included for dimensional reasons, while a dimensionless factor multiplying the right--hand side of Eq. (\ref{ordehache}) is allowed.\footnote{This dimensionless factor is undetermined, in the sense that our argument does not provide its precise value---not in the sense that the viscosity $\eta$ is undetermined.} Moreover, this dimensionless factor will generally depend on the quantum state under consideration, because both $U$ and ${\cal I}$ are state--dependent. Although the viscosity of the quantum probability fluid depends, through an undetermined dimensionless factor, on the quantum state, the order of magnitude provided by Eq. (\ref{ordehache}) is universal.

\subsection{Viscous states {\it vs}\/. dissipation--free states}

Condition (\ref{atlee}) need not be satisfied by all wavefunctions, as the functions $S$ and ${\cal I}$ are already determined by the quantum Hamilton--Jacobi equation and by the continuity equation. Thus our next task is to exhibit a class of quantum--mechanical wavefunctions for which condition (\ref{atlee}) is indeed satisfied, either exactly or at least approximately.

\subsubsection{Exact solutions}

Eq. (\ref{atlee}) integrates to
\begin{equation}
U+\frac{\eta}{\rho}\nabla^2{\cal I}=C_0(t), \qquad C_0(t)\in\mathbb{R},
\label{atleemax}
\end{equation}
where the integration constant $C_0(t)$ may generally depend on the time variable. Let us for simplicity set $C_0(t)=0$. Using (\ref{cupot}) and (\ref{felicita}) the above becomes 
\begin{equation}
\frac{2\eta l^3}{\hbar^2}\nabla^2{\cal I}={\rm e}^{2S}\left[\left(\nabla S\right)^2+\nabla^2S\right].
\label{nouec}
\end{equation}
One can regard (\ref{nouec}) as a Poisson equation $\nabla^2\Phi=\varrho$, where the role of the electric potential $\Phi$ is played by the phase ${\cal I}$ and that of the charge density $\varrho$ is played by the right--hand side of Eq. (\ref{nouec}). The bracketed term, $\left(\nabla S\right)^2+\nabla^2S$, is actually proportional to the Ricci scalar curvature of the conformally flat metric $g_{ij}={\rm e}^{-S({\bf x})}\delta_{ij}$, where $\delta_{ij}$ is the Euclidean metric on $\mathbb{R}^3$.  Eq. (\ref{nouec}) has been dealt with in ref. \cite{PERELMAN}, in connection with the Ricci--flow approach to emergent quantum mechanics; it will also be analysed in a forthcoming publication \cite{NOIFUTURO}. For the moment we will relax the requirement that Eq. (\ref{atlee}) hold exactly, and will satisfy ourselves with approximate solutions instead.

\subsubsection{Approximate solutions}

Under the assumption that $\rho$ is spatially constant, Eq. (\ref{atlee}) integrates to
\begin{equation}
U({\bf x}, t)=C_1(t), \qquad C_1(t)\in\mathbb{R},
\label{izzue}
\end{equation}
where Eqs. (\ref{atled})  and (\ref{laplas}) have been used; the  integration constant $C_1(t)$ may however be time--dependent. Equivalently, one may assume that $S$ in (\ref{nouec}) is approximately constant as a function of the space variables,  hence ${\cal I}$ is an approximate solution of  the Laplace equation (\ref{laplas}). Still another way of arriving at (\ref{izzue}) is to assume the flow to be approximately incompressible, $\nabla\cdot{\bf v}\simeq 0$. Of course, $\rho=mA^2/l^3$ is generally not spatially constant. However, in the semiclassical limit, the amplitude $A={\rm e}^S$ is a slowly--varying function of the space variables. Under these assumptions,  Eq. (\ref{izzue}) holds approximately: 

{\bf Theorem 3} {\sl In the semiclassical limit, the sufficient condition (\ref{atlee}) guaranteeing the validity of the Navier--Stokes equation is equivalent to Eq. (\ref{izzue}).}

We can now consider the effect of taking the semiclassical limit in the identification $(\nabla p)/\rho=(\nabla V)/m$ made in Eq. (\ref{potpre}). In this limit $\rho$ is approximately constant, and the above identification defines an integrable equation for the scalar field $p$. Therefore: 

{\bf Theorem 4} {\sl In the semiclassical limit, the identification $(\nabla p)/\rho=(\nabla V)/m$ made in Eq. (\ref{potpre}) correctly defines a scalar pressure field $p$ within the probability fluid.}

In the stationary case, when $\psi=\phi({\bf x})\exp(-{\rm i}Et/\hbar)$, the quantum potential becomes time--independent, and condition (\ref{izzue}) reduces to the requirement that $U$ be a constant both in space and in time:
\begin{equation}
U({\bf x})=C_2, \qquad C_2\in\mathbb{R}.
\label{izzuezz}
\end{equation}

{\bf Theorem 5} {\sl In the semiclassical limit of stationary eigenfunctions, the sufficient condition (\ref{atlee}) guaranteeing the validity of the Navier--Stokes equation is equivalent to Eq. (\ref{izzuezz}).}

One expects semiclassical stationary states to possess vanishing viscosity because, having a well--defined energy, they are dissipation--free. This expectation is borne out by a simple argument:  Eq. (\ref{atlee}) and the (approximate) spatial constancy of $U$ imply $\eta\nabla^2{\bf v}=0$. This reduces the Navier--Stokes equation (\ref{noxba}) to the Euler equation for a perfect fluid. Therefore:

{\bf Theorem 6} {\sl All semiclassical stationary states have vanishing viscosity: $\eta=0$.}

Thus, as far as dissipation effects are concerned, the combined assumptions of stationarity and semiclassicality lead to a dead end. Furthermore, we cannot lift the requirement of semiclassicality because stationarity alone does not guarantee that the sufficient condition (\ref{atlee}) holds. Even if we per decree assign a non--semiclassical but stationary state $\eta=0$, that state need not satisfy condition (\ref{atlee})---the very assignment of a viscosity $\eta$ would be flawed. 

A physically reasonable assumption to make is that viscosity must be proportional to the density of the fluid:
\begin{equation}
\eta=C_3\rho. 
\label{biscosita}
\end{equation}
Here $C_3$ is some dimensional conversion factor that does not depend on the space variables: $C_3\neq C_3({\bf x})$. Then Eq. (\ref{atlee}) integrates to
\begin{equation}
U+mC_3\left(\nabla\cdot{\bf v}\right)=C_4, \qquad C_4\in\mathbb{R}.
\label{ffee}
\end{equation}
When the flow is incompressible, $\nabla\cdot{\bf v}=0$, and Eq. (\ref{ffee}) reduces to the case already considered in Eqs. (\ref{izzue}) and (\ref{izzuezz}). Thus the proportionality assumption (\ref{biscosita}) provides an independent rationale for the semiclassical approximation made earlier, and viceversa. In turn, this shows that the semiclassicality condition can be recast as done in Eq. (\ref{laplas}). We conclude:

{\bf Theorem 7} {\sl In the semiclassical limit, the viscosity $\eta$ is proportional to the density $\rho$ of the quantum probability fluid. In particular, the viscosity $\eta$ is approximately spatially constant for semiclassical states. Moreover, the proportionality factor $C_3$ in Eq. (\ref{biscosita}) is linear in Planck's constant $\hbar$:}
\begin{equation}
C_3=\frac{\hbar}{m}f.
\label{bzcocho}
\end{equation}
Here $f\geq 0$ is an arbitrary dimensionless factor. By what was said previously, $f=0$ when the state considered is an energy eigenstate, while $f>0$ on all other states.
Hence $f$ is best thought of as a function $f:{\cal H}\rightarrow \mathbb{R}$ on the Hilbert space ${\cal H}$ of quantum states.

Having exhibited the existence of approximate solutions to condition (\ref{atlee}),  whenever dealing with dissipation effects we will restrict our discussion to {\it nonstationary states}\/.

\subsection{The ratio of viscosity to entropy density}\label{razzon}

We have interpreted dissipation as a quantum effect within the probability fluid. Hence the increase ${\rm d}s/{\rm d}t$ in the volume density of entropy of the probability fluid also qualifies as a quantum effect. Here we will compute ${\rm d}s/{\rm d}t$ in the semiclassical regime, both for stationary and nonstationary states.

Considering a stationary state first, we expect ${\rm d}s/{\rm d}t=0$ because $\eta=0$. This expectation is confirmed by the following alternative argument. We see that Eq. (\ref{calor}) reduces to
\begin{equation}
\frac{{\rm d}s}{{\rm d}t}=\frac{\partial s}{\partial t}+({\bf v}\cdot\nabla)s=\frac{\kappa}{\rho }\frac{\nabla^2T}{T},
\label{colar}
\end{equation}
because the dissipation term $\sigma'_{ik}$ vanishes. On the other hand, by Boltzmann's principle (\ref{volman}) we can write the entropy ${\cal S}$ in terms of the amplitude $A={\rm e}^S$ as 
\begin{equation}
{\cal S}=2k_B\ln\left({\Big\vert}\frac{\psi}{\psi_0}{\Big\vert}\right)=2k_B\ln A.
\label{amplitudo}
\end{equation}
This is reminiscent of the expression for the entropy of an ideal gas as a function of its temperature, {\it viz}\/. ${\cal S}=gk_B\ln(T/T_0)$, with $g$ a dimensionless number and $T_0$ some fixed reference temperature. Which suggests identifying the quantum--mechanical amplitude $A$ with the thermodynamical temperature $T$, at least in the absence of friction---as is indeed the case for stationary states and for the ideal gas. So we set
\begin{equation}
A=\frac{T}{T_0}.
\label{identifiko}
\end{equation}
Thus $\nabla^2A=0$ implies $\nabla^2 T=0$. In the semiclassical approximation, $A$ is a slowly--varying function, and one can approximate $\nabla^2 A$ by zero. Thus substituting Eq. (\ref{identifiko}) into Eq. (\ref{colar}), we arrive at a counterpart to theorem 5:

{\bf Theorem 8} {\sl In the semiclassical approximation, the entropy density of any stationary state is constant in time: ${\rm d}s/{\rm d}t=0$.}

Our next task is to obtain an estimate for the order of magnitude of the entropy density $s$. This is readily provided by Eq. (\ref{amplitudo}):

{\bf Theorem 9} {\sl In the semiclassical approximation, the volume density of entropy $s$ of the quantum probability fluid is proportional to Boltzmann's constant:}
\begin{equation}
s=\frac{1}{l^3}O\left(k_B\right).
\label{habemus}
\end{equation}

As already mentioned regarding Eq. (\ref{ordehache}), the denominator $l^3$ has been included for dimensional reasons, and an undetermined, dimensionless factor multiplying the right--hand side is allowed. Finally combining Eqs. (\ref{ordehache}) and (\ref{habemus}) together we can state: 

{\bf Theorem 10} {\sl For the quantum probability fluid in the semiclassical approximation, the order of magnitude of the ratio of viscosity to entropy density is}
\begin{equation}
\frac{\eta}{s}=O\left(\frac{\hbar}{k_B}\right).
\label{papam}
\end{equation}

Again an undetermined, dimensionless factor multiplying the right--hand side is allowed, but the dependence on the length scale $l$ has dropped out.

\subsection{Nonstationary states: emergent reversibility}\label{loewe}

Nonstationary states can be readily constructed as linear combinations of stationary eigenstates with different energy eigenvalues. The ratio $\eta/s$ of the viscosity to the entropy density of a nonstationary state is important for the following reason. Any nonstationary state thermalises to a final equilibrium state. The time required for this transition is of the order of the Boltzmann time $\tau_B$, 
\begin{equation}
\tau_B:=\frac{\hbar}{k_BT}, 
\label{volmantain}
\end{equation}
where $T$ is the temperature of the final equilibrium state \cite{GOLDSTEIN}. In Eq. (\ref{identifiko}) we have related the temperature $T$ to the amplitude $A=\vert\psi_{\rm eq}\vert$ of the equilibrium state wavefunction $\psi_{\rm eq}$. Therefore:

{\bf Theorem 11} {\sl For semiclassical, nonstationary states of the quantum probability fluid, the Boltzmann time is directly proportional to the ratio $\eta/s$ of the viscosity to the entropy density of the initial state, and inversely proportional to the amplitude of the final equilibrium state.}

Out of this analysis there arises a nice picture of the thermalisation process, whereby a nonstationary state decays into a final stationary state. In this picture we have a slow dynamics superimposed on a fast dynamics. The latter corresponds to nonstationary states; the former, to stationary states. Viscous states  correspond to the fast dynamics, while dissipation--free states pertain to the slow dynamics. Time reversibility emerges as a conservation law that applies only to the emergent, slow dynamics.

\subsection{Stationary states: emergent holography}\label{losanomalos}

Turning now our attention to stationary states, let us see how an emergent notion of holography arises naturally in our context. For stationary states we first set $\partial S/\partial t=0$ in the continuity equation (\ref{konntt}), then apply the semiclassicality condition (\ref{laplas}), next divide through by $\hbar$ and finally switch from ${\cal I}$ to $I$ as per Eq. (\ref{noetattio}). This establishes:

{\bf Theorem 12} {\sl For semiclassical stationary states we have
\begin{equation}
\nabla I\cdot\nabla S=O\left(l^{-2}\right).
\label{noumapp}
\end{equation}
For such states, Eqs. (\ref{izzuezz}) and (\ref{noumapp}) are equivalent.}

In the limit $l\to\infty$ we have $\nabla I\cdot\nabla S=0$, and the foliation $I={\rm const}$\footnote{This is abuse of language. Strictly speaking, the equation $I={\rm const}$ defines only one leaf of the foliation. The foliation itself is the union of all the leaves obtained by letting the constant run over the corresponding range.} intersects orthogonally the foliation $S={\rm const}$. That the length scale $l$, in our case of semiclassical stationary states,  can be regarded as being sufficiently large, follows from Eq. (\ref{skala}). Indeed a classical, perfectly localised state around ${\bf x}={\bf x}_0$ carries a wavefunction $\delta({\bf x}-{\bf x}_0)$, the amplitude of which is almost everywhere zero. As this localised state spreads out, ceasing to be perfectly classical, its width can be taken as an inverse measure of its localisation.
In other words, the limit $\hbar\to 0$ is equivalent to the limit $l\to\infty$. Thus neglecting the right--hand side of Eq. (\ref{noumapp}) we arrive at:

{\bf Theorem 13} {\sl Semiclassical stationary states provide two independent foliations of 3--dimensional space by two mutually orthogonal families of 2--dimensional surfaces, respectively defined by $I={\rm const}$ and by $S={\rm const}$.}

The foliation $I={\rm const}$ is well known since the early days of quantum theory.  On the other hand the foliation $S={\rm const}$ was little used in mechanical contexts until the groundbreaking contributions of refs. \cite{FINSTER2, PADMANABHAN, VERLINDE} to the notion of {\it emergent spacetime}\/. Specifically, in ref. \cite{VERLINDE}, isoentropic surfaces $S={\rm const}$ are taken to be holographic screens, while also qualifying as equipotential surfaces $V={\rm const}$ of the gravitational field.  We see immediately that:

{\bf Theorem 14} {\sl Under the above assumptions of stationarity and semiclassicality,\\
{\it i)} the vector field $\nabla I$ is parallel to the foliation $S={\rm const}$;\\
{\it ii)} the vector field $\nabla S$ is parallel to the foliation $I={\rm const}$;\\
{\it iii)} whenever $\nabla I\neq 0\neq\nabla S$, the vector fields $\nabla I$ and $\nabla S$ define an integrable 2--dimensional distribution on $\mathbb{R}^3$.}

The integrability of the distribution defined by the vector fields $\nabla I$ and $\nabla S$ follows from the semiclassicality property $\nabla I\cdot\nabla S=0$. Then Frobenius' theorem guarantees the existence of a family of 2--dimensional integral manifolds for the distribution.\footnote{A purely differential--geometric proof of this statement can be found in ref. \cite{KOBAYASHI}; a related theorem by Liouville, in the context of classical integrability theory, can be found in ref. \cite{ARNOLD}.} Each leaf of this integral foliation, that we denote by $F={\rm const}$, is such that its two tangent vectors $\nabla S$ and $\nabla I$ point in the direction of maximal increase of the corresponding quantities, $S$ and $I$. Therefore:

{\bf Theorem 15} {\sl Under the above assumptions of stationarity and semiclassicality, the foliation $F={\rm const}$ is orthogonal to the two foliations $S={\rm const}$ and $I={\rm const}$ simultaneously.}

According to ref. \cite{VERLINDE}, the leaves $S={\rm const}$ are holographic screens, enclosing that part of space that can be regarded as having emerged. We see that the leaves $I={\rm const}$ play an analogous role with respect to the time variable. Now the wavefunction contains both amplitude and phase. Hence the two foliations $S={\rm const}$ and $I={\rm const}$ must appear on the same footing---as is actually the case.
Taken together, these  facts can be renamed as the {\it holographic property of emergent quantum mechanics}\/. To be precise, this holographic property has been analysed here in the semiclassical regime only; we defer a full analysis until a forthcoming publication \cite{NOIFUTURO}.

\section{Discussion}\label{diskku}

To first order of approximation, any viscous fluid can be characterised by its viscosity coefficients and by its volume density of entropy. In this paper we have obtained an estimate for the order of magnitude of these quantities, in the case of irrotational flow, for the quantum probability fluid. Our analysis makes decisive use of Madelung's factorisation of the quantum wavefunction into amplitude and phase. However, we deviate substantially from Madelung on  the following key issue: {\it Madelung's probability fluid is ideal, while our is viscous}\/. Correspondingly, Madelung's fluid satifies Euler's equation for a perfect fluid, while ours satisfies the Navier--Stokes equation. Consequently, the pressure within the fluid is also different: in Madelung's analysis, pressure is (proportional to) the quantum potential $U$, while our pressure is (proportional to) the external potential $V$  in the Schroedinger equation. In our alternative approach, {\it the quantum potential is responsible for the appearance of viscosity}\/. Thus classical friction in the fluid can be regarded as the origin of quantum effects. Moreover, the dissipation that is inherent to quantum phenomena, under the guise of viscosity in our case, is a nonstationary phenomenon.

By letting the quantum potential account for the viscosity of the probability fluid, our analysis lends support to the emergent paradigm of quantum mechanics: the resulting theory, once dissipation has been taken into account, is no longer classical but quantum. We regard viscosity as the dissipation, or information--loss mechanism, whereby the fluid described by the Navier--Stokes equation (a classical process) becomes the quantum wavefunction satisfying the Schroedinger equation (a quantum process). This mechanism illustrates the statement {\it quantum $=$ classical $+$ dissipation}\/ made in the introductory section.

\vskip.5cm
\noindent
{\bf Acknowledgements} It is a great pleasure to thank F. Finster and R. Gallego Torrom\'e for interesting technical discussions.


\begin{thebibliography}{99}

\bibitem{PERELMAN}
S. Abraham, P. Fern\'andez de C\'ordoba, J.M. Isidro and J.L.G. Santander, {\it A Mechanics for the Ricci Flow}, Int. J. Geom. Meth. Mod. Phys. {\bf 6} (2009) 759, {\tt arXiv:0810.2356 [hep-th]}.

\bibitem{NOI0}
D. Acosta, P. Fern\'andez de C\'ordoba, J.M. Isidro and J.L.G. Santander, {\it An Entropic Picture of Emergent Quantum Mechanics}, Int. J. Geom. Meth. Mod. Phys. {\bf 9} (2012) 1250048, {\tt arXiv:1107.1898 [hep-th]}.

\bibitem{ARNOLD}
V. Arnold, {\it Mathematical Methods of Classical Mechanics}, Springer, Berlin (1989).

\bibitem{BOHM}
D. Bohm, {\it A Suggested Interpretation of the Quantum Theory in Terms of ``Hidden" Variables. I}, Phys. Rev. {\bf 85} (1952) 166. 

\bibitem{CARROLL}
R. Carroll, {\it On the Emergence Theme of Physics}, World Scientific, Singapore (2010).

\bibitem{COUDER}
Y. Couder, S. Proti\`ere, E. Fort and A. Boudaoud, {\it Walking and Orbiting Droplets}, Nature {\bf 437} (2005) 208. 

\bibitem{DEBROGLIE}
L. de Broglie, {\it La Thermodynamique Cach\'ee des Particules}, Ann. Inst. Poincar\'e (A) Physique Th\'eorique {\bf 1} (1964) 1.

\bibitem{ELZE1}
H.-T. Elze, {\it Symmetry Aspects in Emergent Quantum Mechanics}, J. Phys. Conf. Ser. {\bf 171} (2009) 012034.

\bibitem{ELZE2}
H.-T. Elze, {\it Discrete Mechanics, Time Machines and Hybrid Systems}, EPJ Web of Conferences {\bf 58} (2013) 01013, {\tt arXiv:1310.2862 [quant-ph]}.

\bibitem{NOI1}
P. Fern\'andez de C\'ordoba, J.M. Isidro and Milton H. Perea, {\it Emergent Quantum Mechanics as a Thermal Ensemble}, Int. J. Geom. Meth. Mod. Phys. {\bf 11} (2014) 1450068, {\tt arXiv:1304.6295 [math-ph]}.

\bibitem{NOI2}
P. Fern\'andez de C\'ordoba, J.M. Isidro, Milton H. Perea and J. Vazquez Molina, {\it The Irreversible Quantum}, {\tt arXiv:1311.2787 [quant-ph]}.

\bibitem{NOIFUTURO}
P. Fern\'andez de C\'ordoba, J.M. Isidro and J. Vazquez Molina, in preparation.		

\bibitem{FINSTER1}
F. Finster, {\it The Fermionic Projector, Entanglement and the Collapse of the Wavefunction}, J. Phys. Conf. Ser. {\bf 306} (2011) 012024, {\tt arXiv:1011.2162 [quant-ph]}.

\bibitem{FINSTER2}
F.  Finster, A. Grotz and D. Schiefeneder, {\it Causal Fermion Systems: A Quantum Space-Time Emerging from an Action Principle}, Quantum Field Theory and Gravity, Birkh\"auser (2012) 157, {\tt arXiv:1102.2585 [math-ph]}.

\bibitem{GALLEGO1}
R. Gallego Torrom\'e, {\it A Finslerian Version of 't Hooft Deterministic Quantum Models}, J. Math. Phys. {\bf 47} (2006) 072101, {\tt arXiv:math-ph/0501010}.

\bibitem{GALLEGO2}
R. Gallego Torrom\'e, {\it On the Emergence of Quantum Mechanics, Diffeomorphism Invariance and the Weak Equivalence Principle from Deterministic Cartan-Randers Systems}, {\tt  arXiv:1402.5070 [math-ph]}

\bibitem{GOLDSTEIN}
S. Goldstein, T. Hara and H. Tasaki, {\it The Approach to Equilibrium in a Macroscopic Quantum System for a Typical Nonequilibrium Subspace}, {\tt arXiv:1402.3380 [cond-mat.stat-mech]}.

\bibitem{THOOFT0}
G. 't Hooft, {\it Emergent Quantum Mechanics and Emergent Symmetries}, AIP Conf. Proc. {\bf 957} (2007) 154,  {\tt arXiv:0707.4568 [hep-th]}.

\bibitem{THOOFT1}
G. 't Hooft, {\it The Fate of the Quantum},  {\tt arXiv:1308.1007 [quant-ph]}.

\bibitem{THOOFT2}
G. 't Hooft, {\it The Cellular Automaton Interpretation of Quantum Mechanics},  {\tt arXiv:1405.1548 [quant-ph]}.

\bibitem{KOBAYASHI}
S. Kobayashi and K. Nomizu, {\it Foundations of Differential Geometry},  Wiley, New York (1996).

\bibitem{PLASMA}
P. Kovtun, D. Son and A. Starinets, {\it Viscosity in Strongly Interacting Quantum Field Theories from Black Hole Physics}, Phys. Rev. Lett. {\bf 94} (2005) 111601, {\tt arXiv:hep-th/0405231}.

\bibitem{LANDAUQUANTUM}
L. Landau and E. Lifshitz, {\it Quantum Mechanics}, vol. 3 of {\it Course of Theoretical Physics}, Butterworth--Heinemann, Oxford (2000).

\bibitem{LANDAUFLUID}
L. Landau and E. Lifshitz, {\it Fluid Mechanics}, vol. 6 of {\it Course of Theoretical Physics}, Butterworth--Heinemann, Oxford (2000).

\bibitem{LANDAUER}
R. Landauer, {\it Irreversibility and Heat Generation in the Computing Process}, IBM Journal of
Research and Development {\bf 5} (1961) 183.

\bibitem{QCD}
M. Luzum and P. Romatschke, {\it Conformal Relativistic Viscous Hydrodynamics: Applications to RHIC Results at $\sqrt{s_{NN}}$ = 200 GeV},  Phys. Rev. {\bf C78} (2008) 034915, {\tt arXiv:0804.4015 [nucl-th]}.

\bibitem{MADELUNG}
E. Madelung, {\it Quantentheorie in Hydrodynamischer Form}, Z. Phys.  {\bf 40} (1927), 322.

\bibitem{MALDACENA}
J. Maldacena, {\it The Large N limit of Superconformal Field Theories and Supergravity}, Adv. Theor. Math. Phys. {\bf 2} (1998) 231-252, {\tt arXiv:hep-th/9711200}.

\bibitem{MATONE}
M. Matone, {\it `Thermodynamique Cach\'ee des Particules' and the Quantum Potential}, Ann. Fond. Broglie {\bf 37} (2012) 177, 
{\tt arXiv:1111.0270 [hep-ph]}.

\bibitem{ONSAGER}
L. Onsager and S. Machlup, {\it Fluctuations and Irreversible Processes}, Phys. Rev. {\bf 91} (1953) 1505.

\bibitem{PADMANABHAN}
T. Padmanabhan, {\it General Relativity from a Thermodynamic Perspective}, {\tt arXiv:1312.3253 [gr-qc]}.

\bibitem{PADDY}
S. Kolekar and T. Padmanabhan, {\it Indistinguishability of Thermal and Quantum Fluctuations}, {\tt arXiv:1308.6289 [gr-qc]}. 

\bibitem{DISIPACION}
J. Raftery, D. Sadri, S. Schmidt, H. T\"ureci and A. Houck, {\it Observation of a Dissipation--Induced Classical to Quantum Transition}, {\tt arXiv:1312.2963 [quant-ph]}.

\bibitem{VERLINDE}
E. Verlinde, {\it On the Origin of Gravity and the Laws of Newton}, JHEP {\bf 1104} (2011) 029,  {\tt arXiv:1001.0785[hep-th]}.

\bibitem{ZUREK}
W. Zurek, {\it Decoherence, Einselection, and the Quantum Origins of the Classical}, Rev. Mod. Phys. {\bf 75} (2003) 715, {\tt arXiv:quant-ph/0105127}.


\end{thebibliography}
\end{document}